\documentstyle[twocolumn,aps]{revtex}
\begin{document}

\title{ Complexity of Random  Energy Landscapes, Glass
Transition \\and Absolute Value of Spectral Determinant of Random
Matrices }
\author{Yan V. Fyodorov$^{*}$}
\address{Department of Mathematical Sciences, Brunel
University, Uxbridge UB83PH, United Kingdom}

\date{\today}
\maketitle

\begin{abstract}
Finding the mean of the total number $N_{tot}$ of stationary points
for $N-$dimensional random energy landscapes is reduced to
averaging the absolute value of characteristic polynomial of the
corresponding Hessian. For any finite $N$ we provide the exact
solution to the problem for a class of landscapes corresponding to
the "toy model" of manifolds in random environment. For $N\gg 1$
our asymptotic analysis reveals a phase transition at some
critical value $\mu_c$ of a control parameter $\mu$ from a phase
with finite landscape complexity: $N_{tot}\sim e^{N\Sigma},\,\,
\Sigma(\mu<\mu_c)>0$ to the phase with vanishing complexity:
$\Sigma(\mu>\mu_c)=0$. Finally, we discuss a method of dealing
with the modulus of the spectral determinant applicable to a broad
class of problems.

\end{abstract}

Characterising geometry of a complicated landscape described by a
random function ${\cal H}$
of $N$ real variables ${\bf x}=(x_1,...,x_N)$ is an important
problem motivated by numerous applications
in physics, image processing and other fields of applied mathematics
\cite{Adler}.
The simplest, yet non-trivial task
\cite{sea,halp,toy,spinglass,spinglass1}  is to find the mean number of
all stationary points of ${\cal H}$ (minima, maxima and saddles)
in a given domain of the Euclidean space
by investigating the simultaneous stationarity conditions $\partial_k {\cal H}=0$
for all $k=1,...,N$, with
$\partial_k$ standing for the partial derivative
$\frac{\partial}{\partial x_k}$.
In this context the function
\begin{equation}\label{fundef}
{\cal H}=\frac{\mu}{2}\sum_{k=1}^N x_k^2+V(x_1,...,x_N)
\end{equation}
given by the sum of a purely deterministic quadratic piece
characterized by a non-negative parameter $\mu\ge 0$ and of a
random Gaussian function
$V({\bf x})$ attracted considerable interest for
several independent reasons.
For small $N=1,2$ statistics of stationary points of (\ref{fundef})
were investigated long ago in a classical study of specular light
reflection from a random sea surface\cite{sea} and addressed
several times ever since in various physical contexts, see
\cite{halp,toy}. Most frequently one assumes Gaussian part to be
isotropic with zero mean and correlations
depending only on the Euclidean distance $|{\bf x}_1-{\bf x}_2|^2
=\sum_{k=1}^{N}(x_{1k}-x_{2k})^2$ and given by
\begin{equation}\label{cor}
\langle V({\bf x}_1)V({\bf x}_2)\rangle=
N\,f\left[\frac{1}{2N}|{\bf x}_1-{\bf
x}_2|^2 \right],
\end{equation}
with the brackets standing for the ensemble average.

Recently much interest in Eq.(\ref{fundef}) was
boosted by re-interpreting it as the energy functional of a
certain "toy" models describing elastic manifolds propagating in a
random potential, see \cite{MP,CKD} and references therein.
This type of models is known to display a very nontrivial glassy
behaviour at low enough temperatures - an unusual off-equilibrium
relaxation dynamics attributed to a complex structure of their
energy landscape. Although particular dynamical as well as statical
properties may differ substantially for different
functions $f(x)$ (e.g. "long-ranged" vs. "short-range" correlated
potentials, see \cite{CKD}), the very fact of glassy relaxation is
common to all of them.
In fact, the same model admits an alternative
interpretation as a spin-glass, with $x_i$ being looked at as
"soft spins" in a quadratic well interacting via a random
potential $V$\cite{MP}. From this point of view it is most
interesting to concentrate on the limit of large number of "spins"
$N\gg 1$. The experience accumulated from working
with various types of spin-glass models\cite{spinglass}
suggests that for the energy
landscape to be complex enough to induce a glassy behaviour the
total number of stationary points $N_{tot}(\mu)$
should grow exponentially with $N$ as
$N_{tot}(\mu)\sim\exp{N\Sigma(\mu)}$.
The quantity $\Sigma(\mu)>0$
in such a context is natural to call the {\it landscape
complexity}. On the other hand, it is completely clear that the
number of stationary points should tend to $N_{tot}=1$ for very
large $\mu$ when the random part is negligible in comparison with
the deterministic one. In fact, when $N\to \infty$ we will find
that a kind of sharp transition to the phase with vanishing
complexity occurs at some finite critical value $\mu_{c}$, so that
$\Sigma(\mu)=0$ as long as $\mu>\mu_{c}$, whereas $\Sigma(\mu)>0$
for $\mu<\mu_{c}$ and tends to zero quadratically when
$\mu\to\mu_c$. Such a transition is just the glass transition
observed earlier in a framework of a different approach in
\cite{MP,CKD}.

We start with writing the number of stationary points of
${\cal H}$ in any spatial domain
$D$ as $N_{tot}^{(D)}=\int_D  \rho({\bf x}) \, d^N{\bf x}$, with
$\rho({\bf x})$ being the corresponding
density of the stationary points.
The ensemble-averaged value of such a density can be found as
\[
\rho_{av}({\bf x})=\left\langle
|\det{\left(\mu\delta_{k_1,k_2}+\partial^2_{k_1,k_2} V\right)}|
\prod_{k=1}^N\delta(\mu x_k+\partial_k V)\right\rangle
\]
where $\delta(x)$ and $\delta_{mn}$ stand for the Dirac's
$\delta-$ function and Kroneker symbol, respectively.

To evaluate the ensemble average we notice that for the
Gaussian potential $V$ the first derivatives $\partial_k V$
are Gaussian-distributed and are locally statistically independent of the
second derivatives. Representing the $\delta-$functions as Fourier
integrals and exploiting $\langle\partial_{m}
V\partial_{n} V  \rangle=a^2 \delta_{mn}\,,\ a^2=-f^{'}(0)$
one can easily perform the corresponding part of the
averaging and arrive at:
\begin{equation}
\rho_{av}({\bf x})=\frac{1}{[\sqrt{2\pi} a]^N}e^{-\frac{\mu^2\,{\bf x}^2}
{2a^2}}\left\langle |\det(\mu\delta_{k_1,k_2}+H_{k_1k_2}|\right\rangle
\end{equation}
where we introduced the matrix of second
derivatives of the potential $H_{k_1k_2}\equiv \partial^2_{k_1,k_2} V$. Further
changing $H\to -H$ we see that the problem basically
amounts to evaluating the ensemble average of the
absolute value of the characteristic polynomial $\det(\mu I_N-H)$
(a.k.a. spectral
determinant)  of a particular random matrix $H$.
In particular, the total number of stationary points in the whole
space is given by:
\begin{equation}\label{tot}
N_{tot}(\mu)=\frac{1}{\mu^N}\left\langle|\det(\mu I_N-H)|\right\rangle
\end{equation}

Whenever the physical problem necessitates to
deal with the absolute value of the determinant, its presence
considered to be a serious technical challenge, see \cite{Kurchan}
and references therein. In particular, an intensive work and
controversy persists in calculating the so-called {\it
thermodynamic} complexity of the {\it free} energy for the
standard Sherrington-Kirkpatrick model of spin
glasses\cite{spinglass1} or its generalisations\cite{spinglass}.
Several heuristic schemes based on various versions of the replica
trick were proposed in the literature recently to deal with the
problem, see discussion and further references in
\cite{spinglass1}. Despite some important insights, the present
status of the methods is not yet completely satisfactory.

In the present paper we propose two different methods of dealing with
the modulus of determinants, both free from any mathematical uncertainty.
The first method is specific for the problem in hand
and is heavily based on the isotropy of the correlation function
of the random field $V$ in eq.(\ref{cor}).
Exploitation of this fact provides one
with a possibility to employ, after some
manipulations, the
standard methods of the random matrix theory\cite{Mehta}
and find the explicit expression for the number $N_{tot}$ of stationary
points for any spatial dimension $N$.

To follow such a route we notice that the statistical
properties of the potential $V$ result in the following
second-order moments of the entries $H_{ij}\, \,\{(i,j)=1,...,N\}$:
\begin{equation}
\left\langle H_{il}H_{jm}\right\rangle=\frac{J^2}{N}\left[
\delta_{ij}\delta_{lm}+\delta_{im}\delta_{lj}+
\delta_{il}\delta_{jm}\right]
\end{equation}
where we denoted $J^2=f^{''}(0)$.  This allows one to write
down the
density of the joint probability distribution (JPD) of the matrix $H$
explicitly as
\begin{equation}\label{JPD}
{\cal P}(H)dH\propto dH\exp\left\{-\frac{N}{4J^2}
\left[\mbox{Tr}\left(H^2\right)-\frac{1}{N+2}
\left(\mbox{Tr} H\right)^2\right]\right\}
\end{equation}
where $dH=\prod_{1\le i\le j\le N} dH_{ij}$ and the
proportionality constant can be easily found from the normalisation
condition and will be specified
later on. It is evident that such a
JPD is invariant with respect to rotations $H\to O^{-1}HO$ by
orthogonal matrices $O\in O(N)$, but it is nevertheless
different from the standard one typical for the so-called
Gaussian Orthogonal Ensemble (GOE)\cite{Mehta}. However, introducing one extra
Gaussian integration it is in fact straightforward to relate
averaging over the JPD (\ref{JPD}) to that over the standard GOE.
In particular,
\begin{eqnarray}\label{starting}
&&\left\langle|\det{(\mu I_N-H)}|\right\rangle=\\ \nonumber
&&\int_{-\infty}^{\infty}
\frac{dt}{\sqrt{2\pi}}e^{-N\frac{t^2}{2}}
\left\langle|\det{\left[(\mu+Jt)I_N-H_0\right]}|\right\rangle_{GOE}
\end{eqnarray}
where the averaging over $H_0$ is performed with the GOE-type measure:
$dH_0\,C_N\exp\left\{-\frac{N}{4J^2}\mbox{Tr}H_0^2\right\}$, with
$C_N=N^{1/2}/[(2\pi J^2/N)^{N(N+1)/4}2^{N/2}]$
being the relevant normalisation constant.

To evaluate the ensemble averaging in (\ref{starting})
in the most economic way one
 can exploit explicitly the mentioned rotational $O(N)-$invariance
and at the first step in a standard way\cite{Mehta}
reduce the ensemble averaging to the integration over eigenvalues
$\lambda_1,...,\lambda_N$ of the matrix $H_0$. After a
convenient rescaling $\lambda_i\to J \sqrt{2/N}\lambda_i$
the resulting expression acquires the following form:
\begin{eqnarray}
\nonumber
&&\left\langle|\det{\left[(\mu+Jt)I_N-H_0\right]}|\right\rangle_{GOE}
\propto \int_{-\infty}^{\infty}d\lambda_1\ldots \int_{-\infty}^{\infty}
d\lambda_N \\ &\times& \prod^N_{i<j}|\lambda_i-\lambda_j|
\prod_{i=1}^N |\sqrt{N/2}(m+t)-\lambda_i| e^{-\frac{1}{2}\lambda_i^2}
\end{eqnarray}
where we denoted $m=\mu/J$.
One may notice that the above $N-$fold integral can be further
rewritten as a $N+1$ fold integral:
\begin{eqnarray}\nonumber
&&e^{\frac{N}{4}(m+t)^2} \int_{-\infty}^{\infty}d\lambda_1\ldots
\int_{-\infty}^{\infty}
d\lambda_{N+1}\prod_{i=1}^{N+1} e^{-\frac{1}{2}\lambda_i^2}
\\ \nonumber &\times&
\delta\left(\sqrt{N/2}(m+t)-\lambda_{N+1}\right)
\prod^{N+1}_{i<j}|\lambda_i-\lambda_j|
\end{eqnarray}
Such a representation makes it immediately evident that,
in fact, the expectation value
of the modulus of the determinant in question is simply proportional to
the mean spectral density $\nu_{N+1}[m+t]$ (a.k.a one-point
correlation function $R^{(N+1)}_{1}\left[\sqrt{N/2}(m+t)\right]$,
see \cite{Mehta}) of
the same GOE matrix $H_0$ but of enhanced size $(N+1)\times (N+1)$:
\begin{eqnarray}\label{den}
\nonumber &&\left\langle|\det{\left[(\mu+Jt)I_N-H_0\right]}|\right\rangle_{GOE}
\propto e^{\frac{N}{4}(m+t)^2} \nu_{N+1}[(m+t)]\,,\\
&& \nu_N[\lambda]=\frac{1}{N}\left\langle
\mbox{Tr}\, \delta(\lambda\, I_N-H_0)\right\rangle_{GOE}
\end{eqnarray}

The last relation provides the complete solution
of our original problem for any value of $N$, since the one-point function
$R^{(N+1)}_{1}\left[x\right]$  is known in a closed form\cite{Mehta}
for any value of $N$ in terms of the Hermite polynomials $H_k(x)$.
In particular, for any odd integer $N$ we have:
\begin{eqnarray}\label{R}
\nonumber &&\left\langle|\det{\left[(\mu+Jt)I_N-H_0\right]}|
\right\rangle_{GOE}=\frac{J^N}{\sqrt{2\pi}}
\left[\left(\frac{N-1}{2}\right)!\right]\\ \nonumber
&\times&\left( e^{-\frac{x^2}{2}}
\sum_{k=0}^N \frac{1}{2^kk!}
H^2_k(x)+\frac{1}{2^{N+1}N!}H_N(x)\right.\\
&\times&\left.
\int_{-\infty}^\infty e^{-\frac{u^2}{2}}H_{N+1}(u)
\mbox{sign}(x-u) \,du\right)
\end{eqnarray}
where we denoted $x=\sqrt{\frac{N}{2}}(m+t)$ for brevity.
For even integer $N$ one more term arises,
see \cite{Mehta}.

Being interested mainly in extracting the complexity
$\Sigma(\mu)=\lim_{N\to\infty}N^{-1}\ln{N_{tot}(\mu)}$ we have to
perform an asymptotic analysis of equations
(\ref{tot}),(\ref{starting}), (\ref{den})-(\ref{R}). In principle,
one can employ the known large$-N$ asymptotics of the Hermite
polynomials, but we find it more convenient to use an alternative,
well-known representation for the mean eigenvalue density:
\begin{equation}\label{density}
\nu_N[\lambda]=\frac{1}{N\pi}\mbox{Im}\frac{\partial}{\partial \lambda_b}
\left\langle
\frac{\det{(\lambda I_N-H_0)}}{\det{(\lambda_b I_N-H_0)}}\right\rangle
|_{\lambda_b=\lambda}
\end{equation}
The ensemble average of the ratio of the two determinants can be
easily found in the framework of the supersymmetric
approach\cite{Efetov}. Following a variant of this method
we use for our analysis the following integral
representation for the derivative of the ratio of two determinants
featuring in Eq.(\ref{density}), see Eq.(46) in
Ref.\cite{F}:
\begin{eqnarray}\label{in1}
\nu_N[\lambda]\propto \mbox{Re}\int_{-\infty}^{\infty} \frac{dq}{q^2}
e^{-N{\cal L}(q)}G_N(q,\lambda)
\end{eqnarray}
where ${\cal L}(q)=\frac{q^2}{2}+i\lambda q-\ln{(q)}$ and
\begin{eqnarray}\label{in2}
&&G_N(q)=\int_{0}^{\infty}dp_1\frac{(p_1-q)}{p_1^{3/2}}
\exp\{-\frac{N}{2}{\cal L}(p_1)\}\\ \nonumber
&\times&\int_{0}^{\infty}dp_2\frac{(p_2-q)}{p_2^{3/2}}|p_1^2-p_2^2|
\exp\{-\frac{N}{2}{\cal L}(p_2)\}
\end{eqnarray}
The form of the above expressions Eqs.(\ref{in1}-\ref{in2})
suggests that the large-$N$
asymptotics should be given by a saddle-point contribution
in all integration variables.
It turns out however that the situation is not that simple.
To perform the asymptotic analysis
accurately it is convenient first to get rid of the
non-analyticity in the integrand by passing in the last
expression to new variables $0\le r<\infty, -\infty<\theta<\infty$ by
$p_1=re^{\theta}, p_2=re^{-\theta}$. Further introducing
$u=r(\cosh{\theta}-1)$ we arrive at
\begin{eqnarray}\label{in3}
&&G_N(q)\propto\int_{0}^{\infty}\frac{dr}{r^{2}}
\exp\{-N{\cal L}(r)\}\\ \nonumber
&\times&\int_{0}^{\infty}du (r+u)[(r-q)^2-2qu]
e^{-N[u^2+2u(r+i\lambda/2)]}
\end{eqnarray}
The saddle-point value in $u-$variable is obviously $u_s=-(r+i\lambda/2)$,
but it can yield no contribution as long as $\mbox{Re}\,r>0$
along the contour of integration. The analysis reveals that this
is indeed the case for $|\lambda|<2$.
Under that condition the
$u-$integral is dominated by vicinity of its end point $u=0$, whereas
integrals over $r$ and $q$ are instead saddle-point dominated,
the dominant saddle-points being $r_s=(-i\lambda+\sqrt{4-\lambda^2})/2$ and
$q_s=(-i\lambda-\sqrt{4-\lambda^2})/2$. Calculating the corresponding
contribution we arrive, as expected, to the standard semicircular
spectral density $\nu[\lambda]=\frac{1}{2\pi}\sqrt{4-\lambda^2}$.
If, however, the parameter $\lambda$ is such
that $|\lambda|>2$, the situation turns out to be very different.
In that case both saddle-point values
$r_s=i(-\lambda\pm \sqrt{\lambda^2-4})/2$ are purely imaginary,
necessitating a part of the steepest descent contour to be chosen along
the imaginary axis $\mbox{Re} \, r=0$. As a result, an additional
contribution from the saddle-point $u_s$ turns out to be operative.
Although such  contribution is exponentially small in comparison
with one dominated by the vicinity of $u=0$, it is the only one
which survives after taking $\mbox{Re}$ in (\ref{in1}).
Taking into account the saddle-pont $u_s$ induces
 modifications of the relevant exponential term
in $r-$variable, which now becomes $\exp{\{N[\frac{r^2}{2}+\ln{r}-
\frac{\lambda^2}{4}]\}}$ and replaces
the former expression $\exp{\{-N{\cal L}(r)\}}$.
The relevant saddle-point for $r$ then turns out to be  $r_s=-i$
as long as $\lambda>2$ and it results in exponentially small
("instanton")  value for the
spectral density:
\begin{equation}\label{instanton}
\nu[\lambda]\propto \exp\left\{-N\left[\frac{1}{4}\lambda\sqrt{\lambda^2-4}
-\ln{\frac{\lambda+\sqrt{\lambda^2-4}}{2}}\right]\right\}, \quad \lambda>2
\end{equation}
where we only kept factors relevant for calculating
the complexity in the limit of large $N$.

Finally, we employ the relation (\ref{den}) between the mean
spectral density and the expectation value of the modulus of the
spectral determinant for GOE matrices, and substitute the
resulting expression into the integral (\ref{starting}). In the
latter we can again exploit the saddle-point method for asymptotic
analysis. For $0<m<1$ the relevant saddle point is $t_s=m$
satisfying $0<\lambda_s=t_s+m<2$, and validating the use of the
semicircular spectral density $\nu[\lambda_s]=
\frac{1}{2\pi}\sqrt{4-\lambda_s^2}$ in the calculation. This
yields
\begin{eqnarray}\label{fin1}
\left\langle|\det{(\mu I_N-H)}|\right\rangle\propto
e^{\frac{N}{2}\,(m^2-1)}\sqrt{1-m^2},\quad 0<m<1
\end{eqnarray}
For $m>1$, however, it turns out that one has to use
Eq.(\ref{instanton}) for the spectral density.
The corresponding saddle-point value $t_s$ in the $t-$integral
 is given by the solution of the equation
$m=\frac{1}{2}(\lambda_s+\sqrt{\lambda_s^2-4})$ for the variable
$\lambda_s=t_s+m$. The solution is easily found to be simply
$\lambda_s=m+m^{-1}$ (note that $\lambda_s>2$ ensuring consistency of the
procedure) which yields the resulting value for the modulus of the
determinant to be given by
\begin{eqnarray}\label{fin2}
\left\langle|\det{(\mu I_N-H)}|\right\rangle\propto
e^{N\ln{m}},\quad m>1
\end{eqnarray}
Invoking our basic relation Eq.(\ref{tot}) for $N_{tot}$
we see that the landscape
complexity $\Sigma(\mu)$ of the random potential function (\ref{fundef})
is given by:
\begin{eqnarray}\label{complexity}
&&\Sigma(\mu)=\frac{1}{2}\,
\left(\frac{\mu^2}{J^2}-1\right)-\ln{\left(\mu/J\right)},\quad
\mu<\mu_{c}=J\\ &&
\Sigma(\mu)=0,\quad \mu>\mu_{c}=J
\end{eqnarray}
Earlier works referred to the critical value
$\mu_{c}=J$ as, on one hand, signalling the onset of
a nontrivial glassy dynamics\cite{CKD}, and, on the other hand,
corresponding to the point of a breakdown of
the replica-symmetric solution\cite{MP}. Our calculation provides
an independent support to the point of view attributing both phenomena
to extensive number of stationary points in the energy landscape.
At the critical value the complexity
vanishes quadratically: $\Sigma(\mu\to \mu_c)\propto
(\mu_c-\mu)^2/\mu_{c}^2$.

Finally, let us very shortly discuss an alternative,
less model-specific technique
of evaluating the absolute value of the
spectral determinant. It is based on the following
useful identity , see e.g.\cite{note}:
\begin{eqnarray}\label{absreg}
&&|\det{(\mu I_N-H)}|=\\ \nonumber
&&\lim_{\epsilon\to 0}
\frac{[\det{(\mu I_N-H)}]^2}{\sqrt{\det{\left((\mu-i\epsilon) I_N-H\right)}}
\sqrt{\det{\left((\mu+i\epsilon) I_N-H\right)}}}
\end{eqnarray}
valid for any matrix $H$ with purely real
eigenvalues. For the particular case of real symmetric matrices $H$ one
can represent the two factors in the denominator of the right
hand side in terms of the gaussian integrals absolutely convergent
as long as $\epsilon>0$. Further representing
the determinantal factors in the numerator in terms of the Gaussian
integral over anticommuting (Grassmann) variables we thus get a
{\it bona fide} supersymmetric\cite{Efetov}
object to be analysed. Simultaneous presence in the starting
expression both $\mu^{\pm}=\mu\pm i\epsilon$ and $\mu$
 makes the calculation in this case more involved in comparison
with just a simple ratio of two determinats, as in (\ref{density}).
It is nevertheless an important fact that possibility
to perform the ensemble average explicitly exists whenever
matrix entries of $H$ are Gaussian-distributed, not requiring any
matrix invariance or even independence of the matrix entries.
Similar strategy may be even employed
when $H$ is a stochastic differential operator with certain Gaussian
part, as in the notorously difficult case of the random field Ising model
\cite{PS}. For this reason it is natural to hope that
the suggested method may appear of certain utility beyond
the present model, e.g. when discussing free energy landscapes for
spin-glass related problems\cite{spinglass,spinglass1}.

Full account of the procedure will be presented in a separate
publication and here we just quote the result of the ensemble averaging
for any finite value of $N$ (cf. Eq.(\ref{in1})):
\begin{eqnarray}\label{finabs}
&&\left\langle|\det{(\mu I_N-H_0)}|\right\rangle_{GOE}
\propto e^{Nm^2}\\ \nonumber
&\times&\int_{-\infty}^{\infty}\int_{-\infty}^{\infty}
 \frac{dq_1dq_2}{(q_1q_2)^2} (q_1-q_2)^4
e^{-N[{\cal L}(q_1)+{\cal L}(q_2)]}\\ \nonumber
&\times& \int_{0}^{\infty}\,dp_1(p_1+q_1)(p_1+q_2)
\int_{0}^{\infty}dp_2\,(p_2-q_1)(p_2-q_2)\\ \nonumber
&\times&
\frac{(p_1+p_2)}{(p_1p_2)^{3/2}}K_0[N\epsilon(p_1+p_2)]
\exp{-\frac{N}{2}\left[{\cal L}(p_1)+\overline{{\cal L}(p_2)}\right]}
\end{eqnarray}
where $m=\mu/J$, $K_0(z)$ stands for the Macdonald function,
${\cal L}(q)=\frac{q^2}{2}+im q-\ln{(q)}$, and the bar stands for
the complex conjugation. Note, that
 because of the logarithmic divergency of the Macdonald function
at small arguments one should keep $\kappa=N\epsilon$ finite when performing
an accurate asymptotic analysis of this integral, and only then
set $\kappa=0$. The resulting expression
reproduces all the features found by us earlier in the present paper.

In summary, we calculated the mean total number $N_{tot}$ of stationary
points for a $N-$dimensional potential consisting of
a quadratic well of strength $\mu$ and of a
random gaussian piece $V$. In particular,
for $N\to \infty$  we found that the potential is
characterised by finite landscape complexity:
$N_{tot}\sim e^{N\Sigma},\,\, \Sigma>0$ as long as
 $\mu<\mu_c$, and for $\mu\to\mu_c$
the complexity $\Sigma$ vanishes quadratically.
Finally, we discuss a general method of calculating the mean absolute
value of the spectral determinant.

The author is grateful to LPTMS, University Paris-Sud for the kind
hospitality and financial support during several visits in the
process of this research, and to R.Adler, O. Bohigas,  J. Kurchan, M.
Mezard, and D. Gangardt for their interest in the work and useful
comments.

%\begin{thebibliography}{99}

\end{document}